\documentclass{amsart}
\usepackage{bm,amsmath,amsthm,amssymb,amsfonts}
 \newtheorem{thm}{Theorem}
 \newtheorem{lemma}[thm]{Lemma} 
 \newtheorem{prop}[thm]{Proposition} 
 \newtheorem{prop*}{Proposition}
\theoremstyle{remark}
\newtheorem{remark}[thm]{Remark}
\newtheorem{claim}[thm]{Claim}
\newtheorem*{claim*}{Claim}

\newcommand{\calf}{\mathcal{F}}

\newcommand{\bfr}{{\bf R}}

\newcommand{\ket}[1]{| #1 \rangle}
\newcommand{\bra}[1]{\langle #1 |}
\newcommand{\tr}{\text{tr}}

\newcommand{\calh}{\mathcal{H}}
\newcommand{\calp}{\mathcal{P}}

\newcommand{\id}{\mathcal{I}}
\newcommand{\cale}{\mathcal{E}}
\begin{document}
\title{Comparison of Information structures and Completely Positive Maps}
\author{Eran Shmaya}
\address{School of Mathematical Sciences, Tel Aviv University,
Israel}\email{gawain@post.tau.ac.il}
\begin{abstract} A theorem of Blackwell about comparison between information structures
in classical statistics is given an analogue in the quantum
probabilistic setup. The theorem provides an operational
interpretation for trace-preserving completely positive maps, which
are the natural quantum analogue of classical stochastic maps. The
proof of the theorem relies on the separation theorem for convex
sets and on quantum teleportation.
\end{abstract}

\thanks{I am grateful to Ehud Lehrer for introducing me to Blackwell's
theorem and for many hours of discussions. Also I thank the
anonymous referees for their helpful suggestions and comments}
\maketitle
\section{Introduction}\label{intro}
Consider an observer with access to a quantum particle $S$ which
is entangled with another quantum particle $N$. Let $\calh_S$ and
$\calh_N$ be the corresponding Hilbert spaces and $\Phi$ the
density operator over $\calh_N\otimes\calh_S$ that represents the
state of the bipartite system. This paper considers the
information that the observer can garner about $N$ via
measurements over $S$. We call the triple $(\calh_N,\calh_S,\Phi)$
an information structure.

The approach follows Blackwell's similar analysis in classical
statistics (\cite{Blackwell}). The analysis is comparative. Given a
pair of information structures, $(\calh_N,\calh_S,\Phi)$ and
$(\calh_N,\calh_T,\Psi)$, we seek for a condition under which
$(\calh_N,\calh_S,\Phi)$ can be said to be more informative than
$(\calh_N,\calh_T,\Psi)$. The concept of being `more informative' is
understood operationally, that is in terms of the payoffs that the
observer can expect in a certain class of games. We consider the two
scenarios that correspond to these structures: In the first scenario
the observer has access to a particle $S$ such that the joint state
of $N$ and $S$ is $\Phi$, and in the second scenario he has access
to a particle $T$ such that the joint state of $N$ and $T$ is
$\Psi$. In both scenarios, the observer is assumed to be engaged in
some decision problem, or game, in which he has to choose an action
depending on his estimation about the outcome of future measurements
over $N$. We say that the information structure
$(\calh_N,\calh_S,\Phi)$ is better then the information structure
$(\calh_N,\calh_T,\Psi)$ if, for every possible game, the expected
payoff for the observer in the first scenario is at least as good as
the expected payoff in the second scenario.

Before delving into formal definitions, consider the following
example. Assume that all particles have spin-$\frac{1}{2}$, and that
\[\Phi=\frac{1}{2}(\ket{01}\bra{01}-\ket{10}\bra{01}-\ket{01}\bra{10}+\ket{10}\bra{10})\]
is the density matrix of a singlet and that
\[\Psi=\frac{1}{4}(\ket{00}\bra{00}+\ket{01}\bra{01}+\ket{10}\bra{10}+\ket{11}\bra{11})\]
is the density matrix of two independent random states. Suppose
for example that the observer is involved in the following game.
After performing measurements over his particle, he has to guess
the component of the spin of $N$ along axis $\hat{n}$ for some
fixed unit vector $\hat{n}\in\bfr^3$. His payoff $+1$ for a
correct guess and $-1$ for a wrong guess. In the first scenario,
when the observer has access to a particle $S$ such that the
composite system of $N$ and $S$ is at state $\Phi$, he can
guarantee payoff $+1$ by measuring the spin of $S$ along $\hat{n}$
(which is, with probability one, in the opposite direction to the
spin of $N$.) On the other hand, in the second scenario, when the
observer has access to a particle $T$ such that the composite
system of $N$ and $T$ is at state $\Psi$, measuring $T$ will give
him no help, and, whatever strategy he uses for his guess, his
expected payoff is zero. In fact the situation described by $\Phi$
is better than the situation described by $\Psi$, in the sense
that in \emph{every} ``game'' of this type --a formal definition
of game is given below-- the observer can do better (in a weak
sense) in the former situation.

The advantage of $\Phi$ over $\Psi$ is also reflected in the fact
that an observer who has an access to a particle $S$ such that the
composite system of $N$ and $S$ is at state $\Phi$ can perform
physical manipulations on $S$ that transform the state of the
bipartite system to $\Psi$. To do that, he applies $S$ to the
completely depolarizing map given by
\begin{equation}\label{completely-depolarizing}\rho\mapsto\frac{1}{4}\sum_{\mu=0}^3\sigma_\mu\rho\sigma_\mu,
\end{equation}
where
\begin{equation}\label{pauli}
\sigma_0=\bigr(\begin{smallmatrix}1 & 0\\0
&1\end{smallmatrix}\bigl), \sigma_1\bigr(\begin{smallmatrix}0 &
1\\1 & 0\end{smallmatrix}\bigl),
\sigma_2=\bigr(\begin{smallmatrix}0 &-i\\i
&0\end{smallmatrix}\bigl), \sigma_3=\bigr(\begin{smallmatrix}1 &
0\\0&-1\end{smallmatrix}\bigl).
\end{equation}

Essentially, Theorem~\ref{thethm} says that if
$(\calh_N,\calh_S,\Phi)$ is better than $(\calh_N,\calh_T,\Psi)$
in the sense that for all possible games it is better for the
observer to play the game in the scenario corresponding to
$(\calh_N,\calh_S,\Phi)$, then the observer can perform physical
manipulations on $S$ that transform $\Phi$ into $\Psi$.

As a second example, consider a pair of spin-$\frac{1}{2}$ particles
$N$ and $Q$ whose joint state is given by
\[\Upsilon=\Upsilon_{NQ}=\frac{1}{2}(\ket{01}\bra{01}+\ket{10}\bra{10}).\]
This is a separable state, corresponding to a mixing of two pure
product states. Assume that the observer, who as before has an
access to the second particle $Q$, has to guess the spin of the
first particle $N$ along axis $\hat n$ for some unit vector $\hat
n$, with payoffs as before. If $\hat n=\hat z$ (the $z$-axis) then
the observer can he guarantee payoff $+1$ by measuring the spin of
$Q$ along $\hat{z}$ (which is, with probability one, in the
opposite direction to the spin of $N$ along the same axis.) Thus,
in terms of the operational approach taken by this paper for
measuring of information, $\Upsilon$ is strictly better then
$\Psi$. However, if $\hat n=\hat x$ (the $x$-axis), one can verify
that whatever strategy the observer employs, his expected payoff
under $\Upsilon$ would be zero. Therefore $\Phi$ is better than
$\Upsilon$. Note that the observer can transform a pair of
particles at state $\Phi$ to a pair of particles at state
$\Upsilon$ by applying over $S$ the map
\[\rho\mapsto
\frac{1}{2}\sigma_0\rho\sigma_0+\frac{1}{2}\sigma_3\rho\sigma_3,\]
where $\sigma_\mu$ are given in~(\ref{pauli}), and he can
transform a pair of particles at state $\Upsilon$ to $\Psi$ by
applying over $Q$ the completely depolarizing map
(\ref{completely-depolarizing})

Note that in the last example, although $\Upsilon$ and $\Psi$ are
both separable, $\Upsilon$ is in some games strictly better then
$\Psi$ and is always at least as good as $\Psi$. Thus, the
operational comparison of information structures is not just a
comparison of the amount of entanglement between the two parts.
Indeed quantum particles can be correlated without being
entangled, and this correlation should also be taken into account
when comparing information structures.

Finally we remark that the order `better' over information
structures is a partial order: There exists pairs of structures
$\Phi$ and $\Psi$ that are incomparable, that is for some games
$\Phi$ can yield strictly higher payoffs than $\Psi$ and for some
games $\Psi$ can yield strictly higher payoffs than $\Phi$.

\bigskip

Section~\ref{classical} introduces Blackwell's Theorem in
classical statistics. The quantum analogue is given in
section~\ref{quantum} and proved in section~\ref{theproof}.
Section~\ref{quantitative} discusses relationship to quantitative
measure of correlation and Section~\ref{cntrexmple} discusses the
difference between the classical and quantum setup, which is
related to the existence of positive maps which are not completely
positive.
\section{Blackwell's Theorem in Classical
Statistics}\label{classical} A \emph{classical information
structure} is given by a triple $(N,S,p)$ where $N$ and $S$ are
finite sets and $p$ is a (classical) distribution over $N\times
S$, i.e. $p=(p_{\{n,s\}})_{n\in N,s\in S}$ such that $p_{n,s}\geq
0$ and $\sum_{n,s}p_{n,s}=1$. We can think of $N$ and $S$ as the
sets of possible states of two classical particles.

A \emph{game} is given by a finite set $A$ whose elements are
called \emph{actions}, each action $a\in A$ corresponding to a
payoff function $M^a:N\rightarrow \bfr$. The game is played as
follows: First a pair $(n,s)\in N\times S$ is randomly chosen
according to $p$. The observer sees $s$ (the state of particle
$S$) and then chooses an action $a\in A$.  The observer's payoff
is given by $M^a(n)$
We assume that the observer is \emph{rational}, i.e. that he
chooses his action in according to a \emph{strategy} that
maximizes his expected payoff. Formally, a strategy is given by a
partition $\calp=\{P^a\}_{a\in A}$ of the set of signals (that is,
$P_a$ are disjoint subsets of $S$ whose union is $S$.) If the
observer uses that strategy, then, if he sees the signal $s$, he
chooses the action $a$ such that $s\in P^a$. His payoff is given
by \[ \sum_a \sum_n\sum_{s\in P^a} p_{n,s}M^a(n).\] A rational
player chooses a strategy that maximizes this entity. His expected
payoff is thus given by
\[\max_\calp \sum_a \sum_n\sum_{s\in P^a} p_{n,s}M^a(n),\]
where the maximum ranges over all partitions $\calp=\{P^a\}_{a\in
A}$ of $S$.

Let $(N,S,p)$ and $(N,T,q)$ be two information structures.
$(N,S,p)$ is said to be \emph{better} than $(N,T,q)$ if, for every
game (that is, for every finite set $A$ and every payoff functions
$\{M^a:N\rightarrow\bfr\}_{a\in A}$), the expected payoff to the
rational observer if the game is played over $(N,S,p)$ is at least
as good as the expected payoff if the game is played over
$(N,T,q)$. Thus the partial order 'better' over information
structures is defined in terms of games. Blackwell's Theorem
(\cite{Blackwell}, see also \cite{recent} for a recent survey)
characterizes the same order in purely probabilistic terms:

\begin{thm}\label{blassicalthm}
Let $(N,S,p)$ and $(N,T,q)$ be two classical information structures.
Then $(N,S,p)$ is better than $(N,T,q)$ if and only if there exists
a matrix $F=(f_{s,t})_{s\in S,t\in T}$ such that $f_{s,t}\geq 0$ and
$\sum_t f_{s,t}=1$ for every $s\in S$ (i.e. $F$ is a stochastic
matrix) and \begin{equation}\label{blassical}q_{n,t}=\sum_s
p_{n,s}f_{s,t} \text{ for every }n,t.\end{equation}
\end{thm}

Note that every stochastic matrix $F=(f_{s,t})_{s\in S,t\in T}$
corresponds to a linear transformation
$\calf:\bfr^S\rightarrow\bfr^T$ that transforms probability
distributions over $S$ to probability distributions over $T$. We
call $\calf$ a (classical) stochastic map. If we think of $p$ and
$q$ as elements of $\bfr^{N\times S}$ and $\bfr^{N\times T}$ then,
using the natural isomorphisms $\bfr^{N\times S}\leftrightarrow
\bfr^N\otimes \bfr^S$ and $\bfr^{N\times T}\leftrightarrow
\bfr^N\otimes\bfr^T$, ~(\ref{blassical}) can be written equivalently
as \begin{equation}\label{classtensor}q=(\id\otimes
\calf)p,\end{equation} where $\id$ stands for the identity map
$\id:\bfr^N\rightarrow\bfr^N$.

In particular, a necessary condition for $(N,S,p)$ to be better than
$(N,T,q)$ is that $p$ and $q$ induce the same marginal distributions
over $N$. For this reason, classical accounts of Blackwell's Theorem
usually define information structures using the conditional
distribution of $s$ given $n$, and not in terms of the joint
distribution as above. The two formulations are equivalent since, in
classical probability, the joint distribution is uniquely determined
by the marginal distribution of $n$ and the conditional distribution
of $s$ given $n$.

In statistical literature, the set $S$ is viewed as a set of
possible \emph{signals} to the observer. The application of the
stochastic map $\calf$ in (\ref{classtensor}) is interpreted as
\emph{simulation}: If the observer receives a signal $s$ he creates,
or simulates, a new signal $t$ from the set $T$, distributed
according to the $s$-th line of the matrix $F$. If the distribution
of $(n,s)$ was $p$, the simulation process results in a new signal
$t$ such that the joint distribution of $(n,t)$ is $q$. Since
stochastic maps correspond to all the physical manipulation that can
be performed over a classical particles, the physical meaning of the
application of $\id\otimes\calf$ over $p$ is that during the
simulation process the observer performs manipulations only upon his
part of the bipartite system.

\section{Blackwell's Theorem in Quantum Statistics}\label{quantum}
In the quantum probabilistic setup, an information structure is
given by a triple $(\calh_N,\calh_S,\Phi)$, where $\calh_N$ and
$\calh_S$ are two finite dimensional Hilbert Spaces and
$\Phi=\Phi_{NS}$ is a density operator over $\calh_N\otimes\calh_S$,
representing the state of a bipartite system of two particles $N$
and $S$, of which the observer can only access $S$. Slightly abusing
notations, we sometimes refer to the state $\Phi$ as the information
structure in cases where there should be no confusion to which
particle in a pair of particles at state $\Phi$ the observer has
access.

There are two concepts which need clarification before we can
formulate an analogue of Theorem~\ref{blassicalthm} in quantum
probability. First, we have to define the notion of game. Second, we
have to find the appropriate analogue of stochastic maps.

We start with the second task. As mentioned above, a stochastic
matrix corresponds to a linear mapping from classical probability
distributions over one set to classical probability distributions
over another set. The first quantum analogue that comes to mind is
a linear mapping that transforms density operators into density
operators. These are sometimes called positive maps. But there is
a crucial difference between stochastic maps in classical
statistics and positive maps in quantum statistics. Whereas for
every classical stochastic map $\calf$, $\id\otimes \calf$ is also
stochastic, there exist positive maps $\cale$ such that
$\id\otimes\cale$ is not positive (A well-known example is given
by the transpose map.) We say that $\cale$ is completely positive,
if $\id\otimes\cale$ is positive over $\calh'\otimes\calh_S$ for
every $\calh'$ where $\id$ is the identity map over $\calh'$. For
more information about completely positive map see, for
example,~\cite{bible}.

We now turn to the definition of game in the quantum framework.
Consider again the spin guessing game described in the introduction.
It can be thought of as a game with two actions, namely `guess up'
and `guess down'. The observer, after performing measurements over
his particle, has to choose one of these actions. If he chooses the
first action, his payoff is $+1$ if the spin of $N$ is up, and $-1$
otherwise, i.e. the payoff is given by the observable
$\hat{n}\cdot\vec{\sigma}$ measured over $N$, where
$\sigma=(\sigma_1,\sigma_2,\sigma_3)$ are the Pauli matrices. If, on
the other hand, he chooses the second action (guess down) his payoff
is given by the observable $-\hat{n}\cdot\vec{\sigma}$.

Roughly speaking, a \emph{game} is given by a finite set of
actions, each action corresponding to some observable that
determines the observer's payoff should he choose that action. But
in order to achieve the desired result, we need a more general
setting, in which some auxiliary bipartite system
$\calh_A\otimes\calh_B$ at a \emph{fixed} state $\rho_{AB}$ is
introduced as part of the game. The observer can perform his
measurements on $S$ and $A$, and the payoff is determined by
observables over $N$ and $B$. We call the system
$\calh_A\otimes\calh_B$ the \emph{environment}.

By introducing the environment we expand the set of games we look
it. Game theoretically speaking, the environment has a natural
interpretation: It represents random entities that, although
independent of the information structure, can affect the
observer's payoff. In classical statistics, limiting the set of
games to games without environment (as we did in
Section~\ref{classical}) bears no consequences with regard to
comparison of information structures. In quantum statistics,
however, one must look at the larger set of games with environment
in order to get a concept of comparison which is physically
meaningful. The issue is related to the existence of positive maps
which are not completely positive, which have no analogue in
classical statistics. We return to it in Section~\ref{cntrexmple}.

Moving now to formal definitions, let
$(\calh_N,\calh_S,\Phi_{NS})$ be an information structure. A
\emph{game} over this structure is given by
$(\calh_A,\calh_B,\rho_{AB},M^1,\dots,M^k)$ where
$\calh_A,\calh_B$ are finite dimensional Hilbert spaces,
$\rho_{AB}$ is a density operator over $\calh_A\otimes\calh_B$ and
$M^i=M^i_{NB}$ is an hermitian operator over
$\calh_N\otimes\calh_B$ for each $i$ ($1\leq i\leq k$).
$M^1,\dots,M^k$ are called \emph{actions}. Thus, the actions in
the game correspond to observables over $\calh_N\otimes \calh_B$.
The game is played as follows: First, the environment
$\calh_A\otimes\calh_B$ is prepared at state $\rho_{AB}$
independently of the system $\calh_N\otimes\calh_S$. Then the
observer can perform measurements on the particles $S$ and $A$.
Using the information he gathered, he then chooses one action from
the set of available actions $\{M^1,\dots,M^k\}$. The observable
corresponding to that action is then measured, and the numerical
outcome of this measurement is the payoff to the observer in the
game. Note that the observables $M^i$ need not commute, since only
one of them is actually measured. We assume that the observer is
\emph{rational}, i.e, that he chooses his action using a strategy
that maximizes his expected payoff in the game. The following
figure illustrates the role of the particles that are involved in
the game.

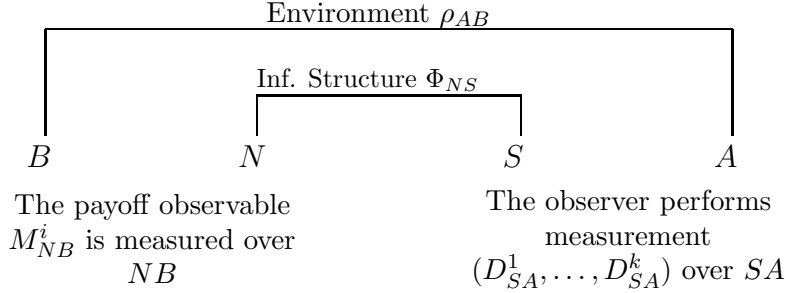
\begin{picture}(300,160)(0,-30)\label{thefigure}\Large
\put(20,30){\makebox{$B$}}\put(100,30){\makebox{$N$}}\put(200,30){\makebox{$S$}}\put(280,30){\makebox{$A$}}
\put(107,42){\line(0,1){15}} \put(107,57){\line(1,0){100}}
\put(207,42){\line(0,1){15}} \normalsize\put(107,60){\makebox{Inf.
Structure $\Phi_{NS}$}}

\put(27,42){\line(0,1){40}} \put(27,82){\line(1,0){260}}
\put(287,42){\line(0,1){40}} \large\put(107,85){\makebox{
Environment $\rho_{AB}$}}

\put(5,0){\parbox[c]{125 pt}{\centering The payoff observable
$M^i_{NB}$ is measured over $NB$}} \put(185,0){\parbox[c]{125
pt}{\centering The observer performs measurement
$(D^1_{SA},\dots,D^k_{SA})$ over $SA$}}
\end{picture}

Note that a strategy in the quantum framework involves the concept
of measurement, which does not appear in the classical framework:
Whereas the classical observer chose his action given the exact
state of his particle, the quantum observer must first choose how
to measure his particle and only then he chooses an action, given
the outcome of the measurement. Formally, a \emph{strategy} is
given by a POVM measurement (\cite{bible}) over $\calh_S\otimes
\calh_A$, i.e. a $k$-tuple
$(D^1,\dots,D^k)=(D^1_{SA},\dots,D^k_{SA})$ of nonnegative
operators over $\calh_S\otimes\calh_A$ such that $D^1+\dots+D^k=I$
(where $I$ is the identity operator). If the observer uses this
strategy, he performs this measurement and chooses action $M^i$ if
the outcome is $i$. His expected payoff is given by
\[\sum_{i=1}^k\tr\bigl((\Phi_{NS}\otimes\rho_{AB})\cdot(M^i_{NB}\otimes
D^i_{SA})\bigr).\] We denote by
$R(\Phi_{NS};\rho_{AB},M^1,\dots,M^k)$ the payoff to the observer
under the best strategy:
\begin{equation}
\label{R_dfen1}R(\Phi_{NS};\rho_{AB},M^1,\dots,M^k)=\max_{(D^1,
\dots,D^k)}\sum_{i=1}^k\tr\bigl((\Phi_{NS}\otimes\rho_{AB})\cdot
(D^i\otimes M^i)\bigr),\end{equation} where the maximum ranges
over all strategies $(D^1,\dots,D^k)$ (that is, over all
$k$-tuples $(D^1,\dots,D^k)$ of nonnegative operators over
$\calh_S\otimes\calh_A$ such that $D^1+\dots+D^k=I$).

\bigskip We now turn to comparison of two information structures. Let
$\Phi_{NS}$ and $\Psi_{NT}$ be two information structures.
$\Phi_{NS}$ is \emph{better} than $\Psi_{NT}$ if, for every
bipartite system $\rho_{AB}$, and every set $\{M^1,\dots,M^k\}$ of
hermitian operators over $\calh_N\otimes\calh_B$, one has
\[R(\Phi_{NS};\rho_{AB},M^1,\dots,M^k)\geq
R(\Psi_{NT};\rho_{AB},M^1,\dots,M^k),\] that is, the observer can
gain in the situation corresponding to $\Phi_{NS}$ at least as much
as he can gain in the situation corresponding to $\Psi_{NT}$. We
prove the following theorem:
\begin{thm}
\label{thethm} Let $\Phi=\Phi_{NS}$ and $\Psi=\Psi_{NT}$ be two
information structures. Then $\Phi$ is better than $\Psi$ if and
only if there exists a completely positive trace preserving map
$\cale_S$ acting on $S$ such that
\begin{equation}\label{thmeq}
\Psi_{NT}=(\id_N\otimes \cale_S)\Phi_{NS},\end{equation} where
$\id_N$ is the identity operation over $N$.
\end{thm}
Note that completely positive trace preserving maps represent the
physical manipulations that the observer can perform on particle
S. Thus the theorem states that $\Phi$ is better than $\Psi$ if
and only if the observer, starting from a pair of particles at
state $\Phi$, can \emph{simulate} a pair of particles at state
$\Psi$ by manipulating only his particle. The `if' part of the
theorem is thus intuitively clear (and easily proved, see
Section~\ref{theproof}): if the observer can achieve some payoff
$r$ in the situation corresponding to $\Psi$, he can achieve the
same payoff in the situation corresponding to $\Phi$. To do that,
he first simulates the situation $\Psi$ by manipulating $S$ and
then applies the strategy that achieves $r$ in situation $\Psi$.
The `only if' part of the theorem says that existence of
trace-preserving completely positive maps that transforms $\Phi$
to $\Psi$ is necessary for the information structure $\Phi$ to be
better than $\Psi$ in the operational sense of allowing higher
payoffs in games.

In particular, it follows from Theorem~\ref{thethm} that a
necessary condition for $\Phi$ to be better than $\Psi$ is that
$\tr_S[\Phi]=\tr_T[\Psi]$, where $\tr_S,\tr_T$ are the partial
traces over $S,T$ respectively. This means that the partial state
of $N$ is the same in both structures. As remarked in
Section~\ref{classical}, there is a similar necessary condition in
the classical framework.

Theorem~\ref{thethm} characterizes the order `better' over
information structures (essentially states) as the order induced by
the set of completely positive, state preserving maps. See Buscemi
et al. \cite{clean} for a related partial order `cleaner' among
POVMs, which is induced by completely positive, identity-preserving
maps.
\section{Proof of Theorem~\ref{thethm}}\label{theproof}
Before proving the theorem, we give another representation of
strategies which is more convenient for the proof. Consider a
strategy in a $k$-action game given by a POVM measurement
$(D^1,\dots,D^k)$. For a density operator $x$ over
$\calh_S\otimes\calh_A$, the probability of getting outcome $i$ if
we perform the measurement $(D^1,\dots,D^k)$ on a particle at state
$x$ is given by
\begin{equation}\label{delta_D}\delta^i(x)=\tr(x\cdot D^i).\end{equation} Note that
$\delta^i$ is a completely positive map from Hermitian operators
over $\calh_S\otimes \calh_A$ to $1\times 1$ matrices. Let
$\delta(x)=(\delta^1(x),\dots,\delta^k(x))$. Then, for every
density operator $x$, $\delta(x)$ is an element of the simplex
$\Delta_k=\{(p_1,\dots,p_k)|p_i\geq 0,p_1+\dots+p_k=1\}$ of
(classical) probability distributions over the set of
actions\footnote{In game theoretic literature elements of the
simplex $\Delta_k$ are called \emph{mixed actions}.}. Thus every
strategy gives rise to a linear function from density operators to
$\Delta_k$. The converse is also true, that is for every linear
function $\delta$ that attaches an element in $\Delta_k$ for every
density operator over $\calh_S\otimes\calh_A$ there corresponds a
POVM measurement $(D^1,\dots,D^k)$ such that (\ref{delta_D}) is
satisfied.

Returning to the game defined by $\rho_{AB}$ and actions
$M^1,\dots,M^k$, recall that the payoff to the observer if he uses
strategy $(D^1,\dots,D^k)$ is
$\sum_{i=1}^k\tr\bigl((\Phi_{NS}\otimes\rho_{AB})\cdot(M^i\otimes
D^i)\bigr)$. Written in terms of $\delta^i$ this amount is given
by $\sum_{i=1}^{
k}\tr\bigl((\id_{NB}\otimes\delta_{SA}^i)(\Phi_{NS}\otimes\rho_{AB})\cdot
M^i\bigr),$ where $\id$ denotes the identity map, and subscripts
of maps correspond to the particle over which they act. Thus
$(\id_{NB}\otimes\delta_{SA}^i)(\Phi_{NS}\otimes\rho_{AB})$ is the
density operator over $\calh_N\otimes\calh_B$ that represents the
joint state of particles $N$ and $B$ after the measurement, if the
outcome was $i$, multiplied by the probability to get outcome $i$.
The maximal possible payoff is given by
\begin{equation}
\label{R_defn}
R(\Phi_{NS};\rho_{AB},M^1,\dots,M^k)=\max_\delta\sum_{i=1}^k\tr\bigl((\id_{NB}\otimes\delta_{SA}^i)
(\Phi_{NS}\otimes\rho_{AB})\cdot M^i\bigr),\end{equation} where the
maximum ranges over all strategies
$\delta=(\delta^1,\dots,\delta^n)$.

We now use this notation to prove the easy `if' part of
Theorem~\ref{thethm}. Let $\Phi_{NS}$ and $\Psi_{NT}$ be two
information structures. Assume that there exists a completely
positive trace preserving map $\cale_S$ acting on $S$ such that
$\Psi_{NT}=(\id_N\otimes \cale_S)\Phi_{NS}$. We prove that
$\Phi_{NS}$ is better than $\Psi_{NT}$. Indeed, let
$\delta=(\delta^1,\dots,\delta^k)$ be a strategy in the $k$-action
game defined by the information structure $\Psi_{NT}$. Thus $\delta$
is a linear function from density operators over
$\calh_T\otimes\calh_A$ to $\Delta_k$. Let $\tilde\delta$ be the
map-composition of $\cale_S\otimes \id_B$ and $\delta$:
$\tilde\delta=\delta\circ(\cale_S \otimes \id_B)$. Since $\cale_S$
is trace preserving and completely positive, it follows that
$\cale_S\otimes \id_B$ is trace preserving and positive, and
therefore $\tilde\delta=(\tilde\delta^1,\dots,\tilde\delta^k)$ is a
strategy in a $k$-action game defined by the information structure
$\Phi_{NS}$.

Now for every environment $\rho_{AB}$ and Hermitian operators
$M^1,\dots,M^k$ over $\calh_N\otimes\calh_B$, since
$\Psi_{NT}=(\id_N\otimes \cale_S)\Phi_{NS}$, it follows that
\begin{equation}
\label{delta_tildelta} (\id_{NB}\otimes
\tilde\delta_{SA}^i)(\Phi_{NS}\otimes\rho_{AB})=(\id_{NB}\otimes\delta^i_{TA})(\Psi_{NT}\otimes\rho_{AB}).
\end{equation}
In particular, it follows from~(\ref{delta_tildelta})
and~(\ref{R_defn}) that
\begin{multline}
\sum_{i=1}^k\tr\bigl((\id_{NB}\otimes\delta_{TA}^i)(\Psi_{NT}\otimes\rho_{AB})\cdot
M^i\bigr)
\\=\sum_{i=1}^k\tr\bigl((\id_{NB}\otimes\tilde\delta_{SA}^i)(\Phi_{NS}\otimes\rho_{AB})\cdot
M^i\bigr) \leq R(\Phi_{NS};\rho_{AB},M^1,\dots,M^k).\end{multline}
And since this is true for every $\delta$, if follows
from~(\ref{R_defn}) that
\begin{equation}\label{it-follows}
R(\Psi_{NT};\rho_{AB},M^1,\dots,M^k)\leq
R(\Phi_{NS};\rho_{AB},M^1,\dots,M^k).\end{equation}
\begin{remark}\label{theremark}Note that in the proof, the only place we used the fact that $\cale_S$ is completely
positive (and not just positive) is to ensure that
$\cale_S\otimes\id_B$ is positive. In particular, if there exist a
positive trace-preserving map $\cale_S$ such that
$\Psi=(\id_N\otimes \cale_S)\Phi$ then (\ref{it-follows}) is still
satisfied for every game with trivial environment (i.e. games in
which $\dim(\calh_B)=1$.) We return to this point in
Section~\ref{cntrexmple}.\end{remark}

\bigskip

Turning to the second (`only if') part of Theorem~\ref{thethm},
let $\Phi=\Phi_{NS}$ and $\Psi=\Psi_{NT}$ be two states such that
$\Phi$ is better than $\Psi$. We construct a completely positive
trace preserving map $\cale_S$ acting on $S$ such that
$\Psi_{NT}=(\id_N\otimes \cale_S)\Phi_{NS}$. The main idea is that
in order to create $T$ from $S$, the observer invokes a pair of
fictitious agents, Alice and Bob. Alice has access to $S$ and she
wants to send Bob the information encoded in the state of system
$T$. To achieve this, they carry the standard teleportation
protocol, only that Alice, instead of measuring the system $T$,
performs an alternative measurement on $S$ with the same effect.
The existence of such an alternative measurement follows from the
fact that $\Phi$ achieves higher payoff than $\Psi$ in every game,
and, in particular, in games whose environment is
the auxiliary state that is used in the quantum teleportation scheme.\\
\textbf{Step 1: Application of Quantum Teleportation}\\ We start
by recalling the quantum teleportation protocol
\cite{teleportation}. Let $\calh_T$ be a finite dimensional
Hilbert space. Assume that Alice wants to send a particle, at a
(possibly mixed) state $x$ that lives in $\calh_T$, to Bob. To do
so, they use a certain bipartite quantum state $\rho_{AB}$ such
that $\calh_T=\calh_B$ of which Alice holds the subsystem
$\calh_A$ and Bob holds the subsystem $\calh_B$. Consider the
state $x_T\otimes \rho_{AB}$. Alice performs on
$\calh_T\otimes\calh_A$ the von-Neumann measurement corresponding
to a certain basis $\ket{\psi^i},\dots,\ket{\psi^k}$. If she gets
outcome $i$, Bob performs a certain unitary operation $U_i$ over
the system $\calh_B$. This sets the system $\calh_B$ in state $x$.

For an Hermitian operator $r$ over $\calh_T\otimes\calh_A$ let
$\delta^i_{TA}(r)=\bra{\psi^i}r\ket{\psi^i}$. For an Hermitian
operator $y$ over $H_B$ we let $\eta^i(y)=U_i^\ast y U_i$. Using
these notations, we can summarize validity of the protocol with the
following equation:
\begin{equation}
\label{tlprtS} (\sum_i \delta^i_{TA}\otimes\eta^i_{B})(x_T\otimes
\rho_{AB})=x_T,
\end{equation}
For every state $x_T$. As usual, subscripts of maps correspond to
the systems on which they act. Note that the state $\rho_{AB}$
that appears in (\ref{tlprtS}) is the specific (maximally
entangled) state that is used in the teleportation protocol. The
magic also operates if the original particle was entangled with
another particle with corresponding Hilbert space $\calh_N$. We
summarize this with the following proposition:
\begin{prop} \label{tlprtprop}There exist a state $\rho_{AB}$, a linear function $\delta=(\delta^1,\dots,\delta^k)$
from density operators over $\calh_T\otimes\calh_A$ to $\Delta_k$
and trace-preserving completely positive maps $\eta^i$ over
$\calh_B$ such that for every density operator $\Psi_{NT}$ over
$\calh_N\otimes \calh_T$ one has
\begin{equation}
\label{tlprtNS}
(\id_N\otimes\sum_i(\delta^i_{TA}\otimes\eta^i_{B}))(\Psi_{NT}\otimes
\rho_{AB})=\Psi_{NT}.
\end{equation}\end{prop}
\noindent\textbf{Step 2 - Application of Separation Theorem}\\The
function $\delta=(\delta^1_{TA},\dots,\delta^k_{TA})$ that appears
in equation~(\ref{tlprtNS}) corresponds to a strategy in a
$k$-action game over the information structure $\Psi_{NT}$ with
environment $\rho_{AB}$. We claim that there exists a strategy
$\tilde\delta_{SA}=(\tilde\delta^1_{SA},\dots,\tilde\delta^k_{SA})$
for the information structure $\Phi_{NS}$ such that, for every
$1\leq i\leq k$,
\begin{equation} \label{sepp}(\id_{NB}\otimes
\tilde\delta^i_{SA})(\Phi_{NS}\otimes\rho_{AB})=(\id_{NB}\otimes
\delta^i_{TA})(\Psi_{NT}\otimes\rho_{AB}),\end{equation} where
$\rho_{AB}$ is the state that appears in
Proposition~\ref{tlprtprop}. Indeed, consider the set $C$ of all
$k$-tuples
\[\bigl((\id_{NB}\otimes
\tilde\delta^1_{SA})(\Phi_{NS}\otimes\rho_{AB}),
\dots,(\id_{NB}\otimes\tilde\delta^k_{SA})(\Phi_{NS}\otimes\rho_{AB})\bigr)\]
for some strategy
$\tilde\delta_{SA}=(\tilde\delta^1_{SA},\dots,\tilde\delta^k_{SA})$.
This is a convex compact set in the linear space of all $k$-tuples
of Hermitian operators over $\calh_N\otimes \calh_B$.

If
$\bigl((\id_{NB}\otimes\delta^1_{TA})(\Psi_{NT}\otimes\rho_{AB}),\dots,(\id_{NB}\otimes\delta^k_{TA})(\Psi_{NT}
\otimes\rho_{AB})\bigr)$ is outside $C$ then by the separation
theorem for convex sets \cite{convex}, there exist a hyperplane that
separates it from $C$, i.e. there exist Hermitian operators
$(M^1,\dots,M^k)$ over $\calh_N\otimes \calh_B$ such that for every
strategy
$\tilde\delta_{SA}=(\tilde\delta^1_{SA},\dots,\tilde\delta^k_{SA})$,
\begin{equation}
\label{suchthat} \sum_i
\tr\bigl((\id_{NB}\otimes\tilde\delta^i_{SA})(\Phi\otimes\rho)\cdot
M^i\bigr) < \sum_i
\tr\bigl((\id_{NB}\otimes\delta^i_{TA})(\Psi\otimes\rho)\cdot
M^i\bigr).\end{equation} Consider the game with set of actions
$\{M^1,\dots,M^k\}$ and environment $\rho_{AB}$.
By~(\ref{suchthat}) and~(\ref{R_defn}) we have
\begin{multline*} R(\Phi_{NS}; \rho_{AB},M^1,\dots,M^k)<\sum_i
\tr\bigl((\id_{NB}\otimes\delta^i_{SA})(\Psi_{NT}\otimes\rho_{AB})\cdot
M^i\bigr)\\\leq R(\Psi_{NT};\rho_{AB},M^1,\dots,M^k),\end{multline*}
contradicting the assumption that $\Phi_{NS}$ is better than
$\Psi_{NT}$. Thus there exists a strategy
$\tilde\delta_{SA}=(\tilde\delta^1_{SA},\dots,\tilde\delta^k_{SA})$
satisfying~(\ref{sepp}).\\
\textbf{Step 3 - Constructing $\cale_S$}\\ It follows
from~(\ref{sepp}) that, for every $1\leq i\leq k$,
\begin{equation}
\label{sepp_eta}
\id_N\otimes\tilde\delta^i_{SA}\otimes\eta^i_B(\Phi_{NS}\otimes\rho_{AB})=
\id_N\otimes\delta^i_{TA}\otimes\eta^i_B(\Psi_{NT}\otimes\rho_{AB}).
\end{equation}
From~(\ref{tlprtNS}) and~(\ref{sepp_eta}) we get
\begin{equation}
\label{finally} \id_N\otimes(\sum_{i}
(\tilde\delta^i_{SA}\otimes\eta^i_{B}))(\Phi_{NS}\otimes
\rho_{AB})=\Psi_{NT}.
\end{equation}
Let $\cale_S$ be the map defined over the system $\calh_S$ by
\begin{equation}
\label{dollar} \cale_S:x_S\mapsto \sum_{i}
(\tilde\delta^i_{SA}\otimes\eta^i_{B})(x_S\otimes
\rho_{AB}).\end{equation} This is a trace preserving, completely
positive map. From~(\ref{finally}) and~(\ref{dollar}) we get
\[
(\id_N\otimes \cale_S)(\Phi_{NS})=\Psi_{NT},\] as desired.

\section{Quantitative Measure of Information}\label{quantitative}
Recall the useful definition of mutual information of a bipartite
state. If $\rho_{XY}$ is a bipartite state, then $I(\rho_{XY})$ is
the real number defined by
\[I(\rho_{XY})=S(\rho_X)+S(\rho_Y)-S(\rho_{XY}),\] where $S(\rho)$
is von-Neumann's entropy of $\rho$. This is the quantum analogue
of mutual information of random variables in classical statistics.
The following lemma is an immediate consequence of
Theorem~\ref{thethm}. Its classical analogue is well known.
\begin{lemma} Let $\Phi=\Phi_{NS}$ and $\Psi=\Psi_{NT}$ be two information structures.
If $\Phi$ is better than $\Psi$ then $I(\Phi)\geq I(\Psi)$.
\end{lemma}
\begin{proof}By Theorem~\ref{thethm}, there exists a trace preserving completely
positive map $\cale$ such that $\Psi=(\id\otimes\cale)\Phi$. Since
local quantum operation cannot increase mutual information, (see,
for example, \cite{bible} Section 11.4.2) it follows that
$I(\Phi)\geq I(\Psi)$.\end{proof} The converse, however, is not true
(neither in classical statistics), as the
following example shows. 

Consider the bipartite state $\Upsilon$ given by
\[\Upsilon=\frac{1}{2}(\ket{01}\bra{01}+\ket{10}\bra{10}).\]
The information structure was already discussed in
Section~\ref{intro}. We consider games in which the observer, who
has access to the second particle, has to guess the spin of the
first particle along the $\hat n$ axis for some $\hat n\in\bfr^3$,
with payoff $+1$ for correct guess and $-1$ for incorrect guess.
The maximal possible payoff is $+1$ if $\hat n=\hat z$ and $0$ if
$\hat n=\hat x$.

Let $\Upsilon'=(H\otimes H)\Upsilon (H\otimes H)$ where
\[H=\frac{1}{\sqrt{2}}\begin{bmatrix}1 & 1\\1 -1\end{bmatrix}.\]
When playing the same games over $\Upsilon'$, the observer maximal
payoff is $+1$ if $\hat n=\hat x$ and $0$ if $\hat n=\hat z$.
Therefore it follows that even though
$I(\Upsilon)=I(\Upsilon')=1$, neither is better than the other in
Blackwell's sense: for some games it is better to play over
$\Upsilon$ and for some games it is better to play over
$\Upsilon'$.
\section{Games with Trivial Environment}\label{cntrexmple}
Theroems~\ref{blassicalthm} and ~\ref{thethm} have a similar
purpose: to provide an operational interpretation for the specific
map. Both uses the concept of game to formulate the operational
aspect of. The main conceptual difference between is the
additional introduction of environment. Two notes are in order.
First, the class of quantum games that was defined in
Section~\ref{quantum} includes as a subset games with trivial
environment, that is games for which
$\dim(\calh_A)=\dim(\calh_B)=1$. This games are the natural
analogue of the classical games that were considered in
Section~\ref{classical}. Similarly, one can expand the set of
classical games to games with classical environment. However, in
the classical framework, the concept of games with environment is
superfluous for comparison of information structures because of
the following claim:
\begin{claim}\label{theclaim}if $(N,S,p)$ and $(N,T,q)$ are two
classical information structures such that the payoff over
$(N,S,p)$ is at least as good as the payoff over $(N,T,q)$ for
every classical game with trivial environment, then the payoff of
$(N,S,p)$ is at least as good as the payoff over $(N,T,q)$ also
for games with non-trivial environment.\end{claim} The claim
follows from Theorem~\ref{blassicalthm} and from the fact that if
$\calf$ is a classical stochastic map then so is $\calf\otimes
\id$. We do not prove it formally here, in order to avoid the
notational encumbrance of classical environment.

In the quantum world, however, there is no analog for
Claim~\ref{theclaim}: There exist information structures $\Phi$
and $\Psi$ such that the payoff over $\Phi$ is at least as good as
the payoff over $\Psi$ for every game with trivial environment,
but not for every game with non-trivial environment. This is why
we had to explicitly define the order relation `better' in
Section~\ref{quantum} using games with environment. The
construction of such a pair $\Phi$ and $\Psi$ of information
structures, which we now describe, is based on the existence of
positive maps which are not completely positive.

Let $\calh_N=\calh_S=\calh_T$ be three $n$-dimensional Hilbert
space. Let $m=n^2$ and $\rho_1,\dots,\rho_m$ a set of $n\times n$
density matrices which are linearly independent in the linear
space of all $n\times n$ Hermitian matrices. Let
$\Phi=\frac{1}{m}(\rho_1\otimes \rho_1+\dots+\rho_m\otimes
\rho_m)$ and $\Psi=\frac{1}{m}(\rho_1\otimes
\rho_1^t+\dots+\rho_m\otimes \rho_m^t)$, where $\rho_j^t$ is the
transpose of $\rho_j$. There exists only one linear map $\cale$
from Hermitian operators to Hermitian operators such that
$\Psi=\id\otimes\cale(\Phi)$, and this map is given by
$\cale(H)=H^t$, which is not completely positive. In particular,
it follows from Theorem~\ref{thethm} that there exists a game
(with no-trivial environment) for which $\Psi$ offers higher
payoff then $\Phi$. On the other hand, since $\cale$ is positive
and $\Psi=\id\otimes\cale(\Phi)$, it follows from the
Remark~\ref{theremark} that for every game with trivial
environment the payoff under $\Phi$ is at least as good as the
payoff under $\Psi$.

\begin{remark}One may conjecture that if $\Phi$ and $\Psi$ are two
structures such that the payoff under $\Phi$ is at least as good
as the payoff over $\Psi$ for every game with trivial environment,
then there exist a positive (but not necessarily completely
positive) trace preserving map $\cale$ such that $\Psi=(\id\otimes
\cale)\Phi$. However, I do not know whether this is
true.\end{remark}
%
%

\end{document}